\documentclass[twocolumn]{aastex631}

\usepackage{xspace, color, comment}
\usepackage{amsmath, amssymb, booktabs}

\newcommand{\asassn}{ASAS-SN\xspace}

\newcommand{\sbunit}{\hbox{\ensuremath{\rm{mag}~\rm{arcsec}^{-2}}\xspace}}

\begin{document}


\title{The ASAS-SN Low Surface Brightness Survey I: Proof-of-Concept and Potential Applications}

\author[0009-0009-4230-8664]{Evan Jennerjahn}
\affiliation{Department of Astronomy, The Ohio State University, 140 West 18th Avenue, Columbus OH 43210}
\affiliation{Center for Cosmology and AstroParticle Physics, The Ohio State University, 191 W. Woodruff Avenue, Columbus OH 43210}

\author[0000-0002-2471-8442]{Michael A. Tucker}
\affiliation{Department of Astronomy, The Ohio State University, 140 West 18th Avenue, Columbus OH 43210}
\affiliation{Center for Cosmology and AstroParticle Physics, The Ohio State University, 191 W. Woodruff Avenue, Columbus OH 43210}

\author[0000-0003-4631-1149]{Benjamin J. Shappee}
\affiliation{Institute for Astronomy, University of Hawai‘i, 2680 Woodlawn Drive, Honolulu, HI 96822, USA}

\author{Christopher S. Kochanek}
\affiliation{Department of Astronomy, The Ohio State University, 140 West 18th Avenue, Columbus OH 43210}
\affiliation{Center for Cosmology and AstroParticle Physics, The Ohio State University, 191 W. Woodruff Avenue, Columbus OH 43210}

\author[0000-0002-1027-0990]{Subo Dong}
\affiliation{Department of Astronomy, School of Physics, Peking University, 5 Yiheyuan Road, Haidian District, Beijing 100871, People's Republic of China}
\affiliation{Kavli Institute of Astronomy and Astrophysics, Peking University, 5 Yiheyuan Road, Haidian District, Beijing 100871, People's Republic of China}
\affiliation{National Astronomical Observatories, Chinese Academy of Science, 20A Datun Road, Chaoyang District, Beijing 100101, China}

\author{Annika H. G. Peter}
\affiliation{Department of Astronomy, The Ohio State University, 140 West 18th Avenue, Columbus OH 43210}
\affiliation{Center for Cosmology and AstroParticle Physics, The Ohio State University, 191 W. Woodruff Avenue, Columbus OH 43210}
\affiliation{Department of Physics, The Ohio State University, 191 W. Woodruff Avenue, Columbus OH 43210}

\author[0000-0003-0943-0026]{Jose L. Prieto}
\affiliation{Instituto de Estudios Astrof\'isicos, Facultad de Ingenier\'ia y Ciencias, Universidad Diego Portales, Avenida Ejercito Libertador 441, Santiago, Chile}
\affiliation{Millennium Institute of Astrophysics MAS, Nuncio Monse\~nor Sotero Sanz 100, Off. 104, Providencia, Santiago, Chile}

\author[0009-0001-1470-8400]{K. Z. Stanek}
\affiliation{Department of Astronomy, The Ohio State University, 140 West 18th Avenue, Columbus OH 43210}
\affiliation{Center for Cosmology and AstroParticle Physics, The Ohio State University, 191 W. Woodruff Avenue, Columbus OH 43210}

\author[0000-0003-2377-9574]{Todd A. Thompson}
\affiliation{Department of Astronomy, The Ohio State University, 140 West 18th Avenue, Columbus OH 43210}
\affiliation{Center for Cosmology and AstroParticle Physics, The Ohio State University, 191 W. Woodruff Avenue, Columbus OH 43210}

\begin{abstract}

The \asassn Low Surface Brightness Survey utilizes the $\sim$7 years of $g$-band CCD data from \asassn (The All-Sky Automated Survey for Supernovae) to create stacked images of the entire sky. It is significantly deeper than previous photographic surveys. Our median/95th percentile cumulative exposure time per field is 58.1/86.8 hours, and our median $3\sigma$ $g$-band surface brightness limit off the Galactic plane ($|b|>20$\textdegree) is 26.1 \sbunit. We image large-scale diffuse structures within the Milky Way, such as multiple degree-spanning supernova remnants and star-forming nebulae, and tidal features of nearby galaxies. To quantify how effective our deep images are, we compare with a catalog of known ultra-diffuse galaxies and find a recovery rate of 82\%. In the future, we intend to use this data set to perform an all-sky search for new nearby dwarf galaxies, create an all-sky Galactic cirrus map, create an all-sky low surface brightness mosaic for public use, and more.

\end{abstract}

\keywords{Dwarf Galaxies --- Low Surface Brightness --- \asassn --- ISM --- Tidal Tails --- Stellar Streams}

 \defcitealias{KandK04}{KK04} 

\section{Introduction} \label{sec:intro}

The low surface brightness (LSB) regime consists of sources fainter than the dark night sky \citep{Bothun}, which is roughly $24$ \sbunit~in the $g$-band. LSB sources probe a broad range of astrophysics. First, we can see large-scale structures within our own Milky Way such as star-forming regions \citep{Carina}, giant molecular clouds \citep{GMCMaybe}, and dust and emission line structures in the interstellar medium \citep[ISM; e.g.,][]{DGLMaybe}. We also see ejected material from novae and supernovae that create diffuse structures on scales of 10s to 100s of parsecs \citep[e.g.,][]{CygSNR}. Outside of our own galaxy, we can see tidal features from galaxy mergers \citep[e.g.,][]{NGC474Shell} and LSB galaxies (LSBGs) such as ultra diffuse galaxies (UDGs) and dwarf galaxies which probe galaxy evolution \citep{reviewEvolve} and dark matter \citep{DarkMatter}.

Despite the usefulness of LSB observations, they are difficult to obtain over large areas. The LSB sky is typically background-limited, so stacking many images is required to push significantly fainter than the sky background. For resolved sources, LSB sensitivity is determined by the total integration time $T$ independent of telescope aperture. One simply coadds $N$ images to reduce the backgrounds, with the depth improving as $T^{-\frac{1}{2}}$ until limited by systematic problems. 


\begin{figure*}
    \centering
    \includegraphics[width=0.48\linewidth]{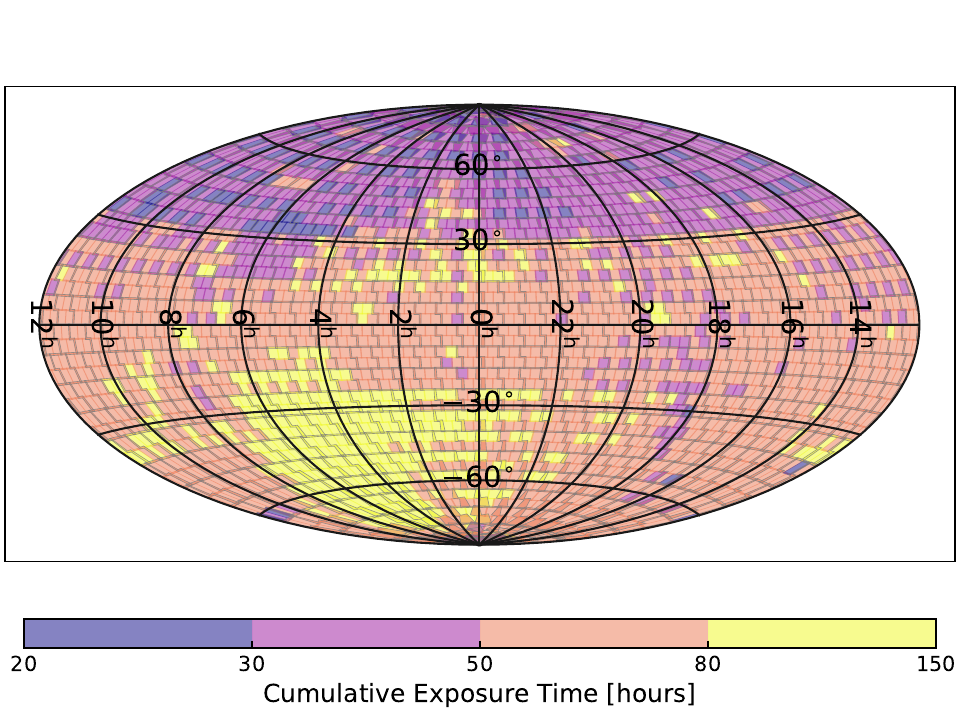}
    \includegraphics[width=0.48\linewidth]{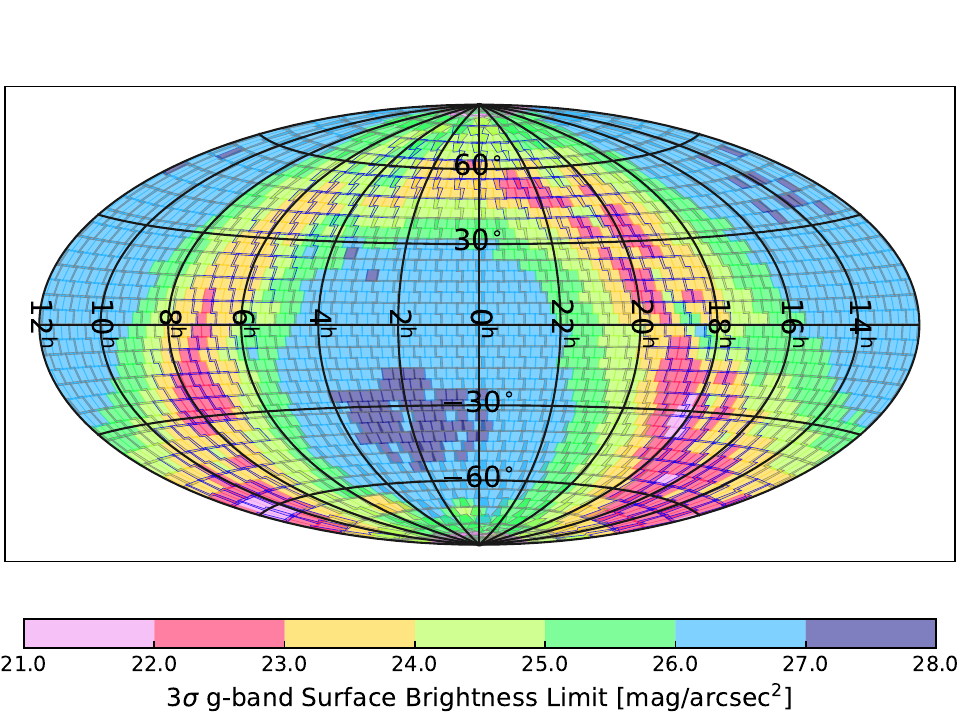}
\caption{Left: The cumulative  $g$-band exposure time combining all cameras of each \asassn field in equatorial coordinates. The Northern hemisphere has 2 telescope mounts while the Southern hemisphere has 3, resulting in the North/South difference. Right: The limiting surface brightness for the best camera for each \asassn field in equatorial coordinates. The structure is due to the high source density in the Galactic Disk and the Magellanic Clouds combined with the low resolution of \asassn.}
     \label{fig:SBLimitMap}
\end{figure*}

Several dedicated projects, such as Huntsman \citep{Huntsman}, Dragonfly \citep{Dragonfly}, and Condor \citep{Condor}, have been built to target the LSB sky. These projects possess several characteristics to maximize their surface brightness sensitivity. Specific design choices include: reducing the number of scattering surfaces, as dust and microroughness backscatter light into the optical path; an unobstructed pupil, because a central obstruction will cause diffraction, resulting in energy being moved to the wings of the PSF; nearly perfect antireflection coatings to prevent ghosts and flaring from polluting the focal plane; and a small focal ratio, because the imaging speed for structures much larger than the resolution limit depends on the focal ratio, not the aperture \citep{OGDragonfly}. An alternative approach to reach LSB depths repurposes archival imaging, typically with custom background filtering or other data processing methods to enhance diffuse structures. For example, imaging from the Legacy Survey \citep{Legacy} was repurposed by several groups to search for low surface brightness galaxies \citep{ Shadows, SmudgeV, EDinLU}. However, each of these programs only cover a fraction of the sky.

\cite{KandK04} \citepalias{KandK04} used digitized photographic plates to make the only true all-sky nearby ($\lesssim 10$ Mpc) galaxy catalog. Their catalog of galaxies includes well-known big, bright galaxies \citep[e.g., M101;][]{M101} and dwarf galaxy candidates \citep{KandK98, Uppsala, Bergh, Lauberts, Paturel}.  Scanned photographic plates are difficult to calibrate \citep{PlateCal1,PlateCal2} and are susceptible to artifacts \citep{PlateArtifacts}. \citetalias{KandK04} also completely avoided the Galactic plane due to crowding. \citetalias{KandK04} make a simple estimate of their completeness, but note that their catalog is likely incomplete. They did not attempt to quantify the limiting depth or angular size of the catalog.

Here we introduce the \asassn Low Surface Brightness Survey which uses years of \asassn \citep{ASASSN, ASASSNLC} imaging, originally intended for time-domain astronomy, to map the entire sky in the $g$-band into the LSB regime using modern CCDs. We provide a description of the data and processing techniques in \S \ref{sec:data}. In \S \ref{sec:apps}, we show that the stacked \asassn images probe a unique discovery space in terms of sky coverage and depth. Finally, \S \ref{sec:summary} summarizes the project and outlines current and future steps to improve our data and fully exploit ASAS-SN's view of the LSB sky.

\section{\asassn Data} \label{sec:data}

\asassn tiles the entire night sky into 2824 overlapping fields and observes them using 20 telescopes on 5 mounts at 4 sites across the globe. In good conditions, \asassn images each of these $\sim4.5\times4.5$ degree fields roughly every night with three 90-second exposures. The survey originally operated in the $V$-band, but switched to the $g$-band in 2017-2018 when it expanded from 2 to 5 mounts. The new units installed in 2017 operated in the $g$-band, and the previous $V$-band mounts switched to $g$-band in 2018. We generated the stacked images in December of 2024.

\begin{figure*}
    \centering
    \includegraphics[width=0.9\linewidth]{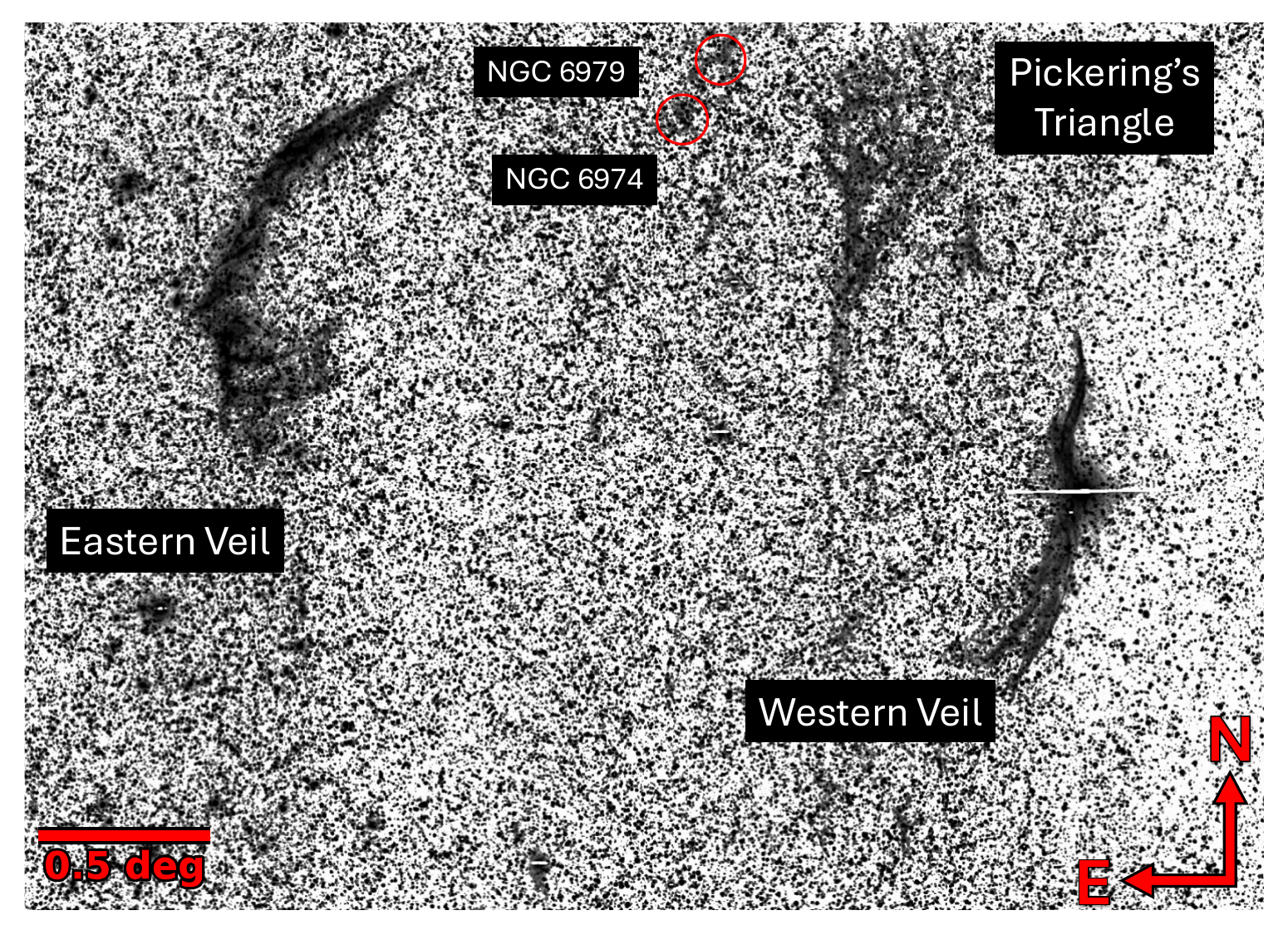}
    \caption{The Veil Nebula, the optical counterpart of the Cygnus Loop supernova remnant in a mosaic made from two stacked \asassn fields. Here, we see both H$\beta$ and [OIII] line emission from the supernova remnant. The radius of the blast wave edge is 1.4 degrees \citep{CygSize}. Several notable features are labeled.}
    \label{fig:VeilNeb}
\end{figure*}

Each image is $2048\times2048$ pixels with a plate scale of $7\farcs8$ per pixel and a median FWHM of 2 pixels ($\sim$16{\arcsec}). The left panel of Fig. \ref{fig:SBLimitMap} shows the cumulative $g$-band exposure time for all \asassn fields. The maximum is 164 hours, with a mean/median of 57.7/58.1 hours. There are three \asassn mounts in the southern hemisphere and only two in the north, resulting in lower total exposure times for $\delta \gtrsim 20^\circ$. The LMC, SMC, and Andromeda are targeted at a higher cadence. Images are vetted to remove those with known issues, including significant cloud coverage, bad flat fields, shutter failure, and poor focus. We then remove the 10\% of the images that have the highest sky backgrounds. After several rounds of trial and error, we found that a 10\% cut gave us the highest quality images with the lowest surface brightless limits.

Each field is observed by at least 2, and up to 5 different cameras leading to the total integration time shown in the left panel of Fig. \ref{fig:SBLimitMap}. Initially, we attempted to combine images across cameras when creating stacked images. However, fields near the poles have significant rotation between images taken by different cameras. The resulting stacks often failed to properly align the images. Thus, for this initial proof-of-concept exercise, we proceed with the single camera stack per field with the deepest surface brightness limit. These limiting surface brightnesses are shown in the right panel of Fig. \ref{fig:SBLimitMap}. The average exposure time for the deepest single camera stack is $35\%$ of what is shown in the left panel of Fig. \ref{fig:SBLimitMap}. However, this varies significantly ($\sim10\%-75\%$) from field to field. In the future, we plan to generate deep images that combine all cameras that observe a given field, which will improve the surface brightness limits.

To combine the images of a field, we first scale all images to the same median pixel value. Mathematically, the median pixel value is a robust estimate of the sky brightness $S_i$ in image $I_i$.  This also means that the noise $\sigma_i$ in the regions without visible sources we will be searching for low surface brightness structures is related to the sky brightness as $\sigma_i^2 \propto S_i$.  Hence, taking the sum $\sum_i I_i/S_i \propto \sum_i I_i/\sigma_i^2$ is equivalent to the standard optimal average of noisy data. Next, we apply iterative pixel-by-pixel $3\sigma$ outlier rejection to produce the final stacked image.

The stacked images are then photometrically calibrated using the \textsc{Refcat} catalog \citep{tonry2018}. The typical point-source sensitivity in a single camera stack is $g\approx 21$~mag which is $\sim 3$~mag deeper than the single-night photometry. Since a single camera stack typically represents only a third of the total field exposure time, the combined camera stacks should have a point source sensitivity of $g\approx 21.75$~mag.

\section{Applications} \label{sec:apps}

\begin{figure*}
    \centering
    \includegraphics[width=0.9\linewidth]{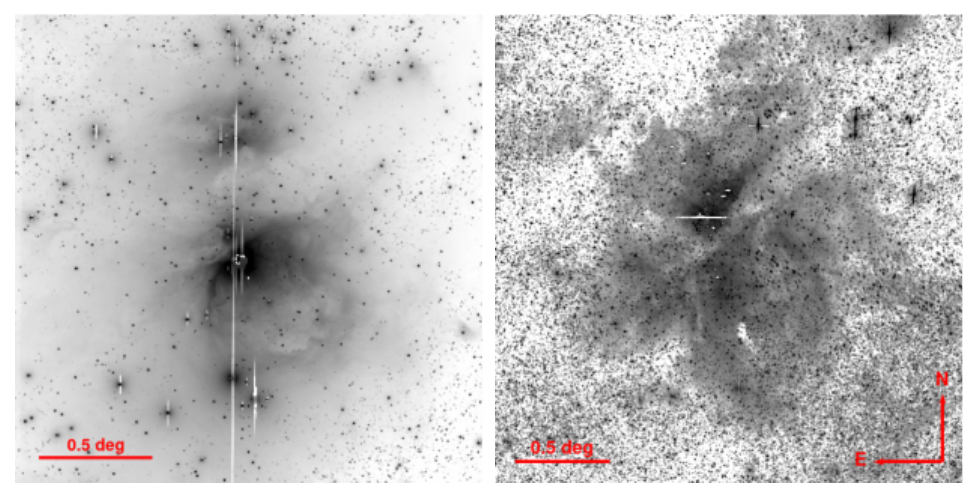}
    \caption{The Orion Nebula (left) and the Carina Nebula (right). The structures are a combination of dust absorption and line emission.}
    \label{fig:2Nebs}
\end{figure*}

\begin{figure}[bt]
    \centering
    \includegraphics[width=0.46\textwidth]{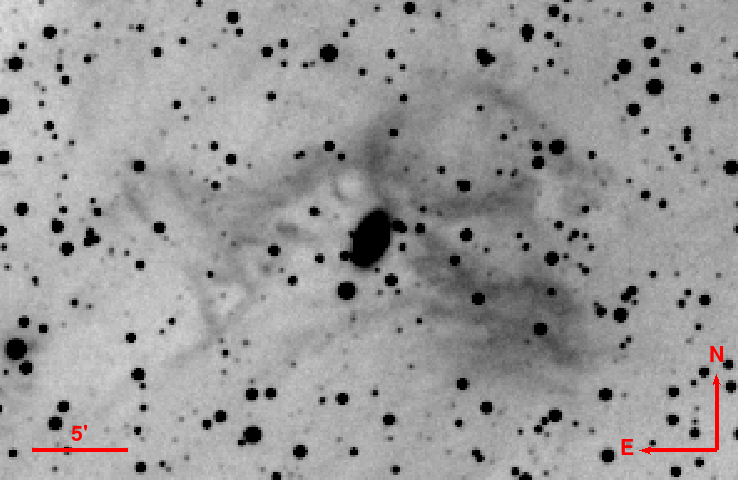}
    \caption{Galactic cirrus around NGC 918 \citep{Amat}.}
    \label{fig:Cirrus}
\end{figure}

The right panel of Fig.~\ref{fig:SBLimitMap} shows the resulting $3\sigma$ surface brightness limits across the sky. Because of our large pixels ($7\farcs8 \times 7\farcs8$), we calculate our surface brightness limit per pixel instead of a more standard $10\arcsec \times 10\arcsec$ or $10~\text{pixel} \times 10~\text{pixel}$ square. The surface brightness limit correlates with stellar density, which is expected given the angular resolution of the \asassn cameras. 2127 of our 2824 total fields (75\%) reach the formal low surface brightness regime ($\gtrsim 24~\sbunit$). This rises to 1805/1859 fields (97\%) off the Galactic plane ($|b|>20$\textdegree).

We illustrate the wide range of use cases for our deep \asassn images using both Galactic (\S\ref{subsec:MW}) and extragalactic (\S\ref{subsec:Tidal} and \S\ref{subsec:lsbg}) examples. We first show our ability to examine features that span several degrees of arc across the sky like the Veil, Carina, and Orion Nebulae. We then focus on our LSB capabilities by examining galactic cirrus, tidal features, and LSBGs. The large pixel scale and wide field-of-view pose unique calibration challenges for identifying smaller sources (${\sim \rm{few}\times\rm{PSF}}$) so we use existing catalogs of LSBGs to better understand our sensitivity on $\sim20\arcsec$ spatial scales in \S\ref{subsec:lsbg}.

\subsection{Objects in the Milky Way}\label{subsec:MW}

\asassn is well suited for looking at large sources spanning several degrees across the sky. Since each \asassn field spans 4.5\textdegree, many large sources can be captured in a single field, or in a mosaic of just a few fields. To demonstrate this, we examined the Veil Nebula, which is the optical counterpart of the $\sim3^\circ$ diameter Cygnus Loop supernova remnant \citep{CygSNR}. Fig. \ref{fig:VeilNeb} shows a mosaic of two stacked fields with 20.4 hours and 20.3 hours of cumulative exposure time, respectively. The $3\sigma$ surface brightness limits of these fields are 23.2 and 23.0 \sbunit. The higher surface brightness limits are due to the proximity of the Veil Nebula to the Galactic plane ($b\approx-10.5$\textdegree). 

\begin{figure*}
    \centering
    \includegraphics[width=0.9\linewidth]{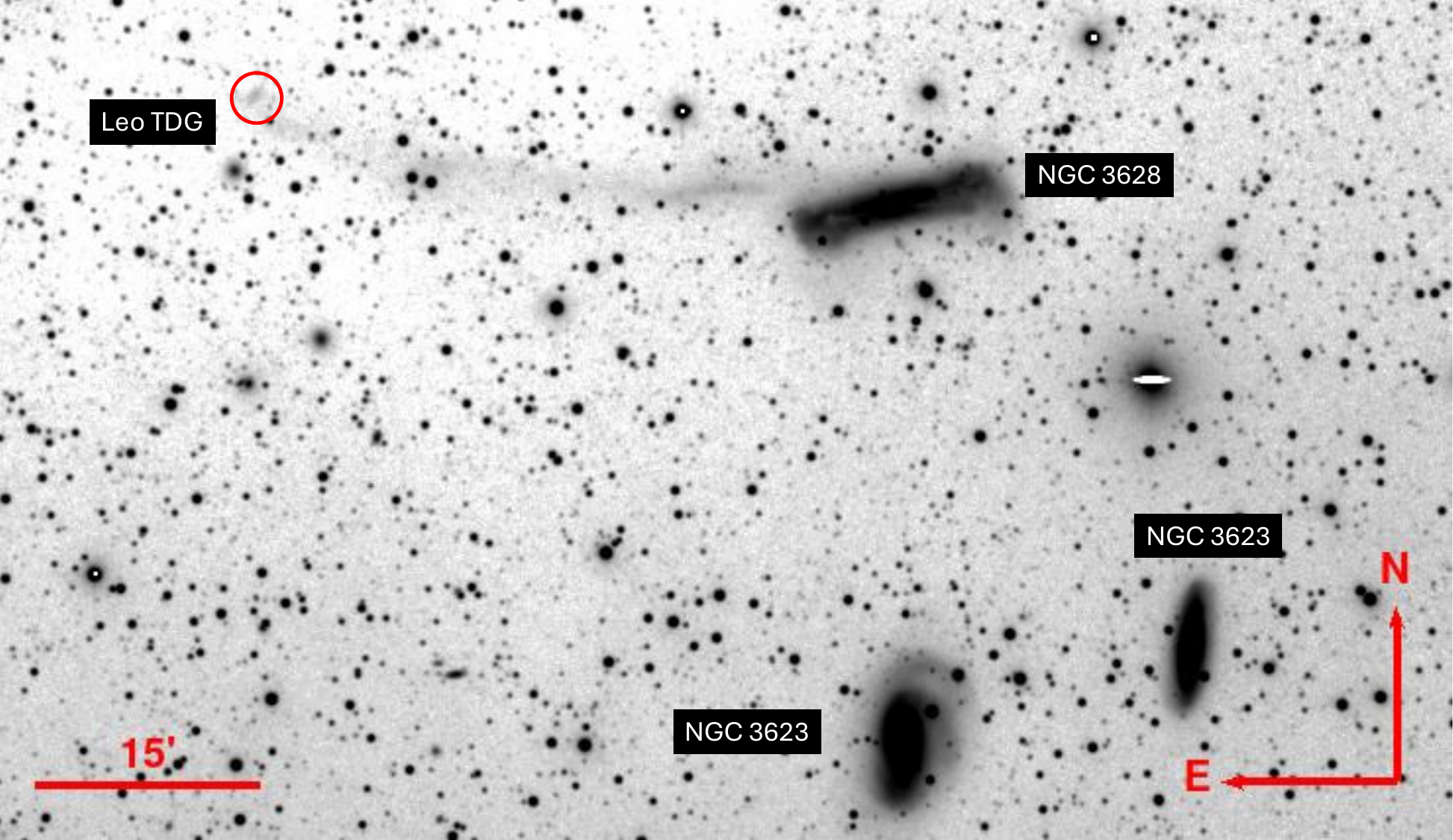}
    \caption{The tidal features around NGC 3628 \citep{OGTail}. The stacked image has a cumulative exposure time of 16.4 hours and a $3\sigma$ surface brightness limit of 26.67. The Leo tidal dwarf galaxy (TDG) can be seen at the tip of the tidal tail \citep{TDG}.}
    \label{fig:TideTail}
\end{figure*}

\begin{figure}[bt]
    \centering
    \includegraphics[width=0.46\textwidth]{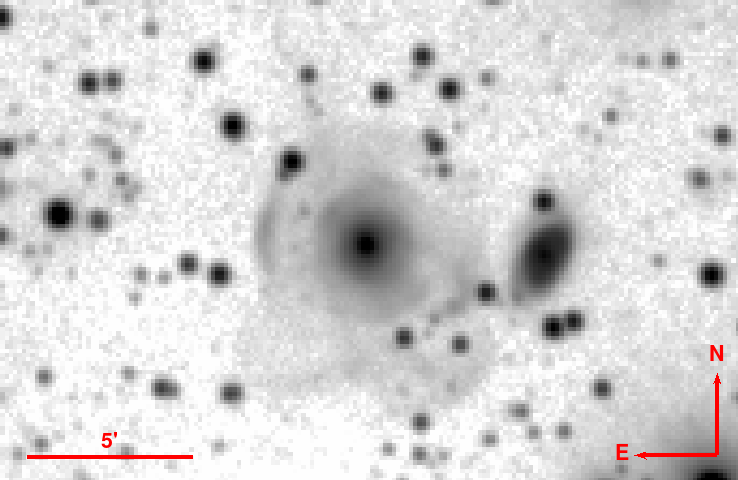}
    \caption{Shell galaxy NGC 474 \citep{Older474} and its neighbor NGC 470 in \asassn. The shells likely formed due to a merger.}
    \label{fig:Shell}
\end{figure}

Fig. \ref{fig:2Nebs} shows images of both the Orion \citep{Orion} and Carina \citep{Carina} nebulae. These nebulae are active star-forming regions within the Milky Way that show diffuse structure from their gas and dust. The Orion Nebula is a mosaic of two stacks with surface brightness limits of 25.25 and 25.71 \sbunit. The Carina Nebula image is a mosaic of three deep stacks with surface brightness limits of 21.76, 21.69, and 21.91 \sbunit. The Carina Nebula mosaic has a much higher surface brightness limit due to the high stellar densities ($b\approx-1$\textdegree), while the Orion Nebula is much farther from the Galactic plane ($b\approx-20$\textdegree).

Lastly, Fig. \ref{fig:Cirrus} shows a deep \asassn image of the Galactic cirrus surrounding NGC 918 \citep{Amat}. Galactic cirrus can cause a range of problems for LSB surveys \citep{Amat}. For example, if the cirrus occupies a large fraction of the field of view, it is difficult to model the image background. There are methods to circumvent this issue \citep{Cirrus}, but they only work for areas of high contamination and on scales similar to the resolution of the Planck and IRAS maps \citep{Planck, IRAS}. The cirrus can also be confused for extragalactic sources or LSB features like stellar streams. Visual identification can be an effective way to recognize cirrus, but it is a subjective process.

\subsection{Tidal Features}\label{subsec:Tidal}

Another example of an LSB structure detectable with stacked \asassn images is merger-induced tidal features. Gravitational interactions between galaxies can produce tidal features that have significant impacts on galactic evolution \citep{OlderMerge, MergersMatter}. Fig. \ref{fig:TideTail} shows NGC 3628 \citep{OGTail} in the Leo Group. There is a long LSB tidal tail formed as a result of NGC 3628 interacting with its companions NGC 3623 and NGC 3627 \citep{StarinTail}. In addition to the tail, there are lobes of faint light extending from both sides of the disk of NGC 3628 down towards the other members of the group. This may be better classified as intragroup light (IGL, see \citealp{DragonEdge}).

\begin{figure*}
    \centering
    \includegraphics[width=0.8\textwidth]
    {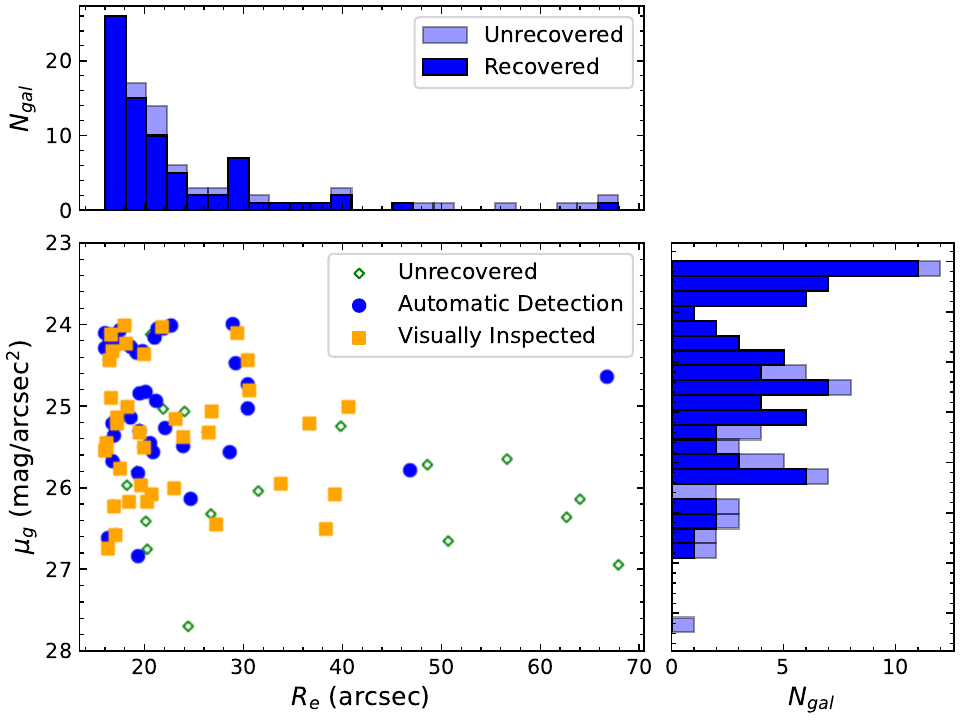}
    \caption{Recovery of SMUDGes LSBGs  in the space of $R_{\rm eff}$ and $\mu_g$ (central surface brightness). We detect 37/92 automatically (filled blue circles), and find an additional 38 through visual inspection (filled orange squares). 17 are undetected (unfilled green diamonds). Many of the unrecovered SMUDGes are unsurprising, as their surface brightness was below our calculated surface brightness limit for their fields.}
    \label{fig:Hist}
\end{figure*}

\begin{figure*}[bt]
    \centering
    \includegraphics[width=0.9\textwidth]
    {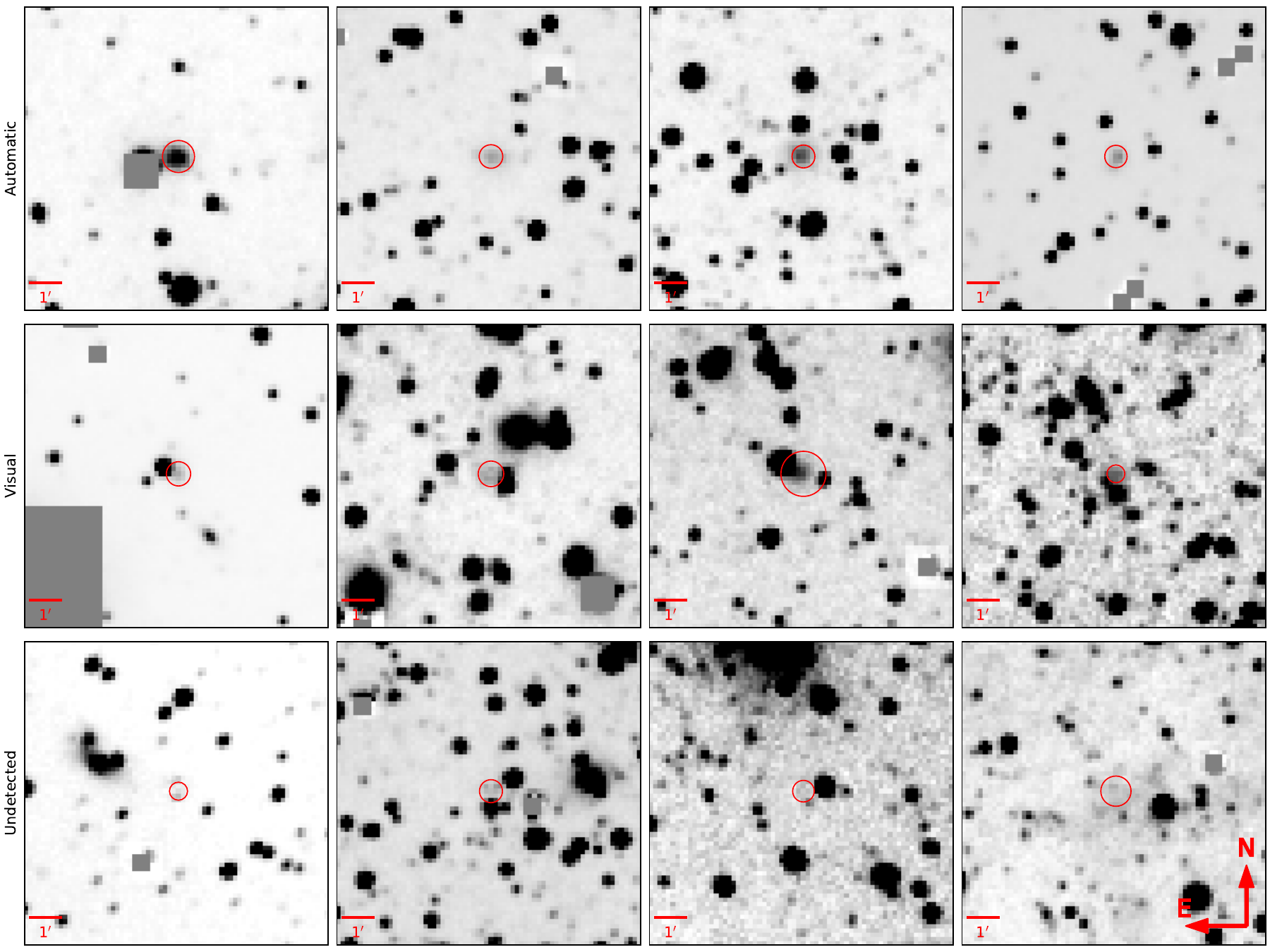}
    \caption{Examples of LSBGs in the stacked \asassn images. The top row displays the LSBGs that were automatically detected by our algorithm. The middle row shows LSBGs that were missed by our algorithm, but were found by visual inspection. The bottom row shows LSBGs that were missed entirely. The ones missed by our recovery algorithm but seen when visually inspected were missed due to the proximity of a brighter source. If an LSBG was entirely missed, it was due to crowding, lying close to the edge of an \asassn field where the flat fields are worse, or a combination of the two.}
    \label{fig:VIExamples}
\end{figure*}

Fig. \ref{fig:Shell} displays LSB features around the shell galaxy NGC 474 \citep{Older474}. Shells, like other faint tidal features, are indicative of hierarchical assembly of galaxies and likely form during galaxy mergers or interactions with nearby galaxies \citep{MergerShell1, MergerShell2, MergerShell3}. The shells formed around NGC 474 likely formed during an intermediate or major merger roughly 2 Gyr ago \citep{NGC474Shell}. Since then, it has been accreting cold gas from its neighbor NGC 470 \citep{NGC474Shell}. Since we are capable of detecting these LSB features around galaxies, a future project will be to image the halos of galaxies and generate their $g$-band surface brightness profiles similar to \citet{DragonEdge}.

\subsection{Low Surface Brightness Galaxies}\label{subsec:lsbg}

Fig. \ref{fig:SBLimitMap} shows that \asassn is capable of reaching surface brightness limits that are competitive with higher-resolution surveys \citep{Shadows, EDinLU, SmudgeV}, especially when looking off the Galactic plane. However, \asassn's lower resolution complicates the identification of LSBGs. To assess our ability to detect and characterize LSBGs, we used the SMUDGes \citep{SmudgeV} catalog of Ultra-Diffuse Galaxies (UDGs) as a benchmark.

In order to mitigate stellar contamination and crowding, we produced a modified image of each field in the DES footprint designed to mask or model and subtract stars. We first mask all stars brighter than 12th magnitude, near the \asassn saturation limit, and stars with proper motions exceeding 1{\arcsec} per year ($\sim$half of the PSF FWHM over the 10-year baseline). Then, we fit and subtract the PSFs of stars between $12-16$ mag with $\geq 5\sigma$ combined proper motion and parallax measurements from \emph{Gaia} \citep{GAIADR3} to remove many Galactic sources. 

To subtract the stars, we first spatially group them to account for blending. Stars are considered to be in the same group if their centers are within 11 pixels of each other. We then create a cutout of each group. Cutouts are $11\times11$ pixels for groups containing only one star, and vary in size for groups containing multiple stars. We then fit each star within a group using a combination of Gaussian and Moffat profiles plus a constant background. Then we subtract the model of the stars, but not the background in order to preserve any LSB structure. The central $2\times2$ pixels of each star are then masked from further analyses because there are still significant residuals compared to the background. This PSF subtraction works well for minimizing the contribution from the stellar PSF wings. Generating these PSF subtracted images is computationally expensive, so we performed an initial test for finding LSBGs using only the fields that overlap with the Dark Energy Survey (DES, \citealp{DESSurvey}). This includes a large portion of the SMUDGes footprint.

After PSF subtraction, we used SourceExtractor \citep{Sextractor} to identify sources within each PSF subtracted field and crossmatched with SMUDGes. The results are shown in Fig. \ref{fig:Hist}. We do not expect to recover LSBGs with an effective radius less than the FWHM of \asassn ($\sim16\arcsec$). Our surface brightness sensitivity drops for unresolved sources and they are increasingly indistinguishable from point sources. Therefore, we only consider LSBGs with $R_{\rm eff} \geq16\arcsec$ .  

Visual inspection is still an important tool when searching for LSBGs \citep{Tucana}. Fig. \ref{fig:Hist} shows that Source Extractor detects some (37/92) of the LSBGs automatically, but we can find 38 more (75/92) using visual inspection. Fig. \ref{fig:VIExamples} shows examples of automatically detected LSBGs (top row), LSBGs that were only detected with visual inspection (middle row), and LSBGs that were not detected (bottom row). The ones only detected with visual inspection are often located in the vicinity of a brighter source, or are significantly larger than the $3\times3$ convolutional kernel we used for SourceExtractor. PSF subtraction helps in these cases (for example, the top left panel of Fig. \ref{fig:VIExamples}) by minimizing light from the nearby source. The objects missed by both methods are either significantly overlapped by nearby stars, or close to the edge of an \asassn field where imperfect flat-fielding causes issues. There are several cases (e.g., the bottom right panel of Fig. \ref{fig:VIExamples}) where there is clearly a source around the UDG location. We do not count these as detections, because when we compare with the higher-resolution DES data, we find that there are several other sources within the red circle, and we cannot definitively say that the signal we see is coming from the UDG.

Thus the \asassn LSB project is sensitive enough to uncover new LSBGs off the Galactic plane. For example, in a conservative estimate, we could detect a Large Magellanic Cloud (LMC)-like object ($M_B=-17.93$, $\mu_B=23.39$ \sbunit, $R=4.98$ kpc; \citealp{LVG,ThirdGal}) out to a distance of $\sim45$ Mpc and a Small Magellanic Cloud (SMC)-like object ($M_B=-16.5$, $\mu_B=24.06$ \sbunit, $R=2.91$ kpc; \citealp{LVG,ThirdGal}) at $\sim20$ Mpc.




\section{Summary and Next Steps}\label{sec:summary}

We have taken the first steps in repurposing \asassn data into a new sensitive all-sky atlas. By stacking $\sim 7$ years of $g$-band observations, we increase the depth of \asassn by $\sim2-4$ magnitudes when compared to single-night exposures. These deep images offer a new opportunity to investigate LSB sources across the entire sky.

Our initial assessments show that we are able to see Milky Way structures spanning several degrees on the sky such as the Veil, Carina, and Orion Nebulae in great detail (\S \ref{subsec:MW}, see Figs. \ref{fig:VeilNeb} and \ref{fig:2Nebs}). Additionally, we recover tidal features around other galaxies such as the long tidal stream around NGC 3628 in the Leo triplet (\S \ref{subsec:Tidal}). Finally, we show that our new deep images are well-suited for an all-sky LSBG search that complements previous searches over smaller areas of the sky (\S \ref{subsec:lsbg}). Through a combination of automatic recovery and visual inspection, we are able to identify 82\% of the UDGs with $R_{\rm eff} \geq 16\arcsec$ in the SMUDGes catalog and within the DES footprint.

\asassn was originally designed for an entirely different science goal than what we pursue here, and this paper serves as our proof-of-concept, in which we showed the ability of stacked \asassn data to probe the LSB regime. We can improve our data even further. The first improvement we can make is combining the data from different cameras to create stacked images utilizing our full cumulative exposure time. This will allow us to see even deeper into the LSB regime. The existing flat fields are not accurate at the $\approx 1\%$ level and produce gradients across our stacked images. This makes it more difficult to clearly identify LSB features near the borders of our fields. However, preliminary testing shows that dividing by a simple 2D polynomial fit helps to remove degree-scale gradients in the stacks. Another area for improvement is our LSBG detection process. Here we only used a $3\times3$ pixel convolutional kernel when running Source Extractor. We should do better when looking for larger LSBGs using a larger kernel. Moreover, there are techniques we could employ to further improve our stacking algorithm \citep[e.g.,][]{Processing}. We have already begun work to improve our PSF subtraction algorithm as well, making it faster, and subtracting more stars.

After making these improvements, there are several possible projects we plan to pursue. We will conduct a search for new LSB galaxies. Our large PSF makes us better suited for searching for larger ($R_{\rm eff} \gtrsim16\arcsec$) LSBGs where many previous studies have focused on smaller ones. We also have the unique advantage of having access to the entire sky to look for LSBGs. Another project is to produce a Galactic cirrus map of the entire sky. Mitigating Galactic cirrus is a concern for the upcoming Rubin survey \citep{Rubin} and we can assist by using our data to map out where known regions of cirrus exist and create a mask of areas to avoid looking for LSB features. Additionally, we can use our data to investigate the LSB halos of galaxies and generate their surface brightness profiles. The deep stacks can also be used to search for light echoes from historical Milky Way supernovae \citep{Echo}. The PSF of \asassn is well matched to existing sky surveys such as GALEX in the ultraviolet \citep{GALEX} and WISE in the infrared \citep{WISE}, providing an easy avenue for multi-wavelength synergy. The end result of this project will be a full-sky $g$-band LSB imaging atlas that will be publicly available and accessible to the community.

\section*{Acknowledgments}
EJ, CSK, and KZS are supported by NSF grants AST-2307385 and 2407206. 

We thank Las Cumbres Observatory and its staff for their continued support of \asassn. \asassn is funded by Gordon and Betty Moore Foundation grants GBMF5490 and GBMF10501 and the Alfred P. Sloan Foundation grant G-2021-14192. 

The Shappee group at the University of Hawai‘i is supported with funds from NSF (grants AST-2407205) and NASA (grants HST-GO-17087, 80NSSC24K0521, 80NSSC24K0490, 80NSSC23K1431).

SD is supported by the National Natural Science Foundation of China (Grant No. 12133005).  S.D. acknowledges the New Cornerstone Science Foundation through the XPLORER PRIZE.

AHGP is supported by National Science Foundation Grant No. AST-2008110.

JLP acknowledges support from ANID, Millennium Science Initiative, AIM23-0001.

This research made use of the cross-match service provided by CDS, Strasbourg. Additionally, we made frequent use of the "Source Extractor for Dummies" user manual written by Dr. Benne W. Howlerda.


%

\vspace{5mm}
\facilities{\asassn}


\software{astropy, lmfit, pyraf, imstat, Source Extractor}







\bibliography{sample631}{}
\bibliographystyle{aasjournal}



\end{document}